\documentclass[%
reprint,
amsmath,amssymb,
aps,
]{revtex4-1}

\usepackage{graphicx}% Include figure files
\usepackage{dcolumn}% Align table columns on decimal point
\usepackage{bm}% bold math
\usepackage{hyperref}% add hypertext capabilities
\usepackage{braket}
\begin{document}

\title{Accessing Rydberg-dressed interactions using many-body Ramsey dynamics}% Force line breaks with \\

\author{Rick Mukherjee}

\email{rm47@rice.edu}

\author{Thomas C. Killian}
\author{Kaden R. A. Hazzard}

\affiliation{Department of Physics and Astronomy, Rice University, Houston, Texas 77005, USA}
\affiliation{Rice Center for Quantum Materials, Rice University, Houston, Texas 77005, USA}

\date{\today}% It is always \today, today,
	
\begin{abstract}
We demonstrate that Ramsey spectroscopy can be used to observe Rydberg-dressed interactions. In contrast to many prior proposals, our scheme operates comfortably within experimentally measured lifetimes, and accesses a regime where quantum superpositions are crucial. We build a spin-1/2 from one level that is Rydberg-dressed and another that is not. These levels may be hyperfine or long-lived electronic states.  An Ising spin model governs the Ramsey dynamics, for which we derive an exact solution. Due to the structure of Rydberg interactions, the dynamics differs significantly from that in other spin systems. As one example, spin echo can \textit{increase} the rate at which coherence decays. The results also apply to bare (undressed) Rydberg states as a special case, for which we quantitatively reproduce recent ultrafast experiments without fitting. 		
	
\begin{description}
	\item[PACS numbers] 32.80Ee, 67.85.-d, 37.10.Jk
\end{description}

\end{abstract}

\pacs{Valid PACS appear here}% PACS, the Physics and Astronomy
\maketitle
	
%\tableofcontents

{\it Introduction} - Ultracold Rydberg atoms allow one to process quantum information ~\cite{Jaksch, Lukin, Gaeten, Urban, Isenhower,Wilk,Saffman}, study interacting many-body systems ~\cite{Weimer0, Olmos0, ovsky1, ovsky2,Pohl0, Viteau3,Hofmann,Carr, ovsky4,Henkel, Cinti, Pupillo}, and engineer nonlinear quantum optics ~\cite{Dudin1,Dudin2, Pritchard, Weatherill, Mohapatra, Petrosyan, Parigi,Peyronel, Gorshkov, Firstenberg, Chang, Maxwell, Li}. These applications stem from the enormous van der Waals interactions between Rydberg atoms excited to large principal quantum number $n\sim 50-70$. Since these interactions are proportional to  $n^{11}$, they are enhanced by many orders of magnitude compared to ground state atoms~\cite{Bohlouli, Beguin}. These interactions inhibit the simultaneous excitation of neighbouring atoms to Rydberg states; this is known as the ``blockade effect"~\cite{Tong, Singer, Vogt, Heideman, Comparat}. Experiments have observed dramatic consequences of Rydberg interactions, such as the formation of Rydberg crystals~\cite{Schauss1,Schauss2} and suppressed excitation number fluctuations in the blockade region~\cite{Liebisch, Viteau, Malossi}. However, observing quantum correlations remains an outstanding challenge. Quantum correlations require superpositions, which in turn require non-commuting terms in the governing Hamiltonian. Obtaining such a situation is difficult with pure Rydbergs, since their interaction energy often dominates the other scales.

%However most of the observations in this regime can be described classically: strong interactions spatially pattern the Rydberg excitations, but the consequences of \textit{superposition} between Rydberg and ground states are negligible.  Fundamentally, this classicality stems from the dominant strength of the Rydberg interaction, as creating quantum superpositions requires non-commuting terms in the governing Hamiltonian. 

Consequently, in order to explore the many interesting applications where superpositions and quantum correlations  are essential~\cite{Henkel, Cinti, Pupillo,Maucher, Wuester, Schempp,Mattioli,Mukherjee,Keating,Gil, Bouchoule, Dalmonte, Bijnen, Glaetzle, dauphin1, dauphin2}, it is necessary to controllably reduce the interaction strength -- for example, one may seek to make it comparable to the Rabi drive in a blockade experiment or comparable to motional energy scales in a trap or lattice.  Dressed Rydberg atoms, in which a small Rydberg fraction is superposed with the ground state, enable this control while simultaneously extending the system lifetime by reducing spontaneous emission~\cite{Honer, Johnson, Balewski}.   An exciting recent breakthrough is the measurement of dressed Rydberg interactions between two atoms~\cite{Jau}, but observing these interactions in a many-body system remains a major outstanding goal. 

\begin{figure}[h!]
	\centering
	\includegraphics[width=\columnwidth]{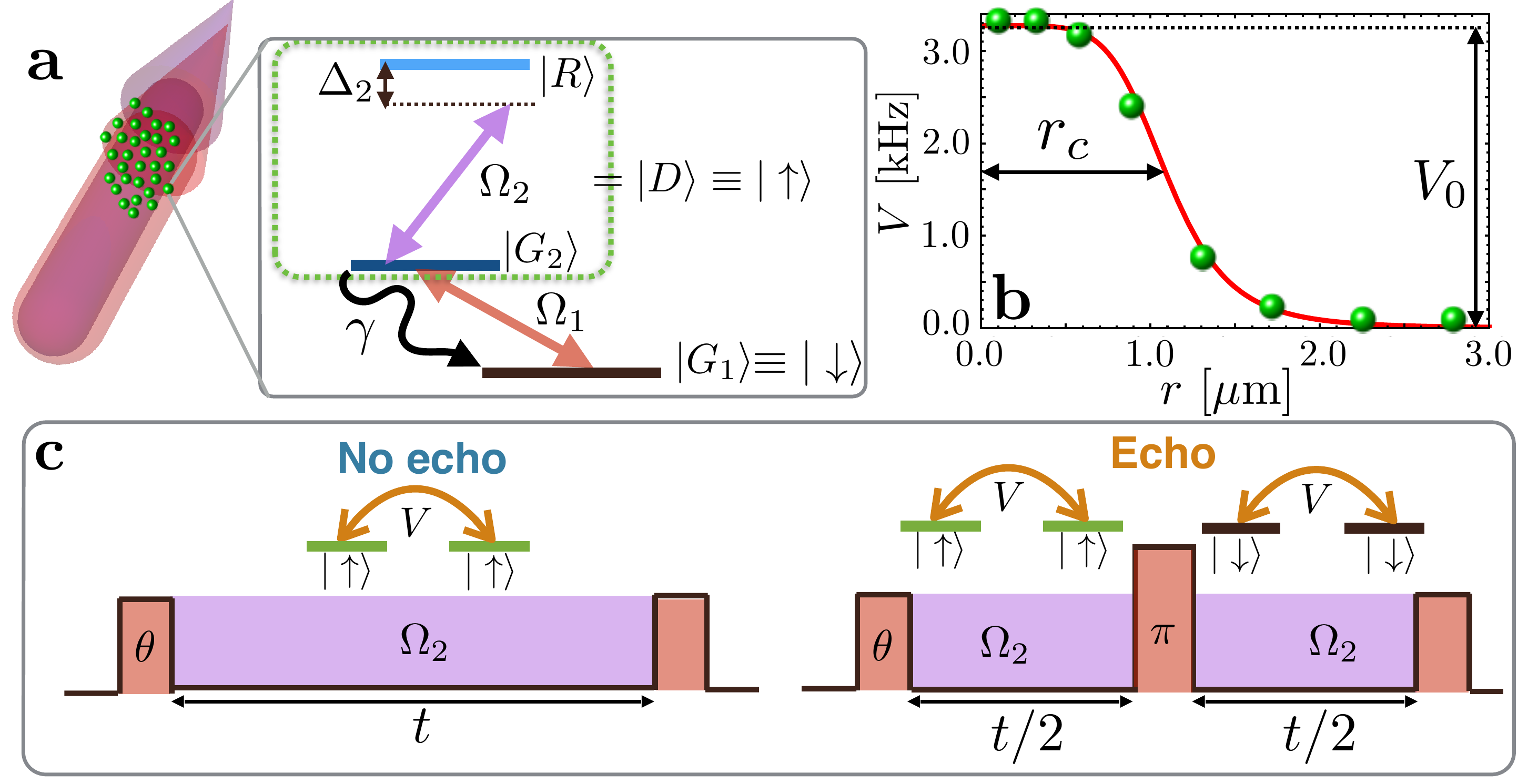}
	\caption{(color online) Probing dressed Rydberg interactions with Ramsey spectroscopy. (a) Atoms (green balls) are dressed and probed with two lasers. One laser ($\Omega_2$, purple) weakly mixes the Rydberg state $|R\rangle$ into $|G_2\rangle$ to form the dressed state $|D\rangle$. This state is probed with strong pulses ($\Omega_1$, red) that couple $|G_1\rangle$ and $|G_2\rangle$. (b) The Rydberg-dressed interaction for the $\Delta_2/C_6<0$ case considered herein is a soft-core potential. (c) Ramsey scheme: An initial $\Omega_1$ pulse with area $\theta$ superposes $\ket{\downarrow} = |G_1\rangle$ and $\ket{\uparrow} = |D\rangle$ states. Dressed atoms interact  between pulses. The final pulse allows one to measure the spin vector. One can apply a $\pi$ echo pulse to eliminate single particle imperfections. This also effectively ``turns on" the interactions between atoms that were initially in the ground state for the second half of the dynamics.}
	\label{setupfig}
\end{figure}	

In this paper, we show how current experiments can observe and characterize dressed Rydberg interactions by using Ramsey spectroscopy, where one probes a spin-1/2 created from two long-lived atomic states, one of which is Rydberg dressed. This Ramsey protocol directly accesses a regime where superpositions are crucial and can give rise to non-classical correlations. Although similar to the Ramsey protocol that has been fruitfully applied to other many-body atomic and molecular systems~\cite{Yan, Zhang2, hazzard0, Hazzard, Trotzky, Koschorreck, Wall, Knap, Rey}, we find that the Rydberg atoms behave in qualitatively different ways because of the shape of the potential and because the ground-Rydberg and ground-ground interactions are negligible compared to the Rydberg-Rydberg interactions. The unique character of the dressed Rydberg interaction also manifests itself in the dependence of the dynamics on atom density, dressed state excitation fraction, spin echo, and  Rydberg state. Just one example of this distinct behavior is that the contrast decay is \textit{faster} with an echo than without in several regimes, including low density, large dissipation, and small excitation fraction of dressed states.

%Our results utilize recently developed exact analytic solutions of the model's dynamics \cite{foss1,foss2}, which incorporate interactions and spontaneous emission. We extend these to analytically average over the random atom positions. 

{\it Setup: Rydberg dressing, interactions, and spectroscopy -} We consider an atom with two long-lived levels $|G_1\rangle $ and $|G_2\rangle$ as shown in Fig.~\ref{setupfig}(a). These could, for example, be two hyperfine states or the ground state and another sufficiently long-lived electronic state. Examples of long-lived electronic states occur in the alkaline-earth-like atoms: the $^3$P$_{0}$ or $^3$P$_1$ states have lifetimes of $159$ s and 21 $\mu$s respectively. We consider the same coupling scheme as in Ref. \cite{Gil}. A laser with Rabi frequency $\Omega_2$ and detuning $\Delta_2$ from resonance admixes a small fraction of the Rydberg state $|R\rangle$ to the $|G_2\rangle$ level, creating a Rydberg-dressed state $|D\rangle\approx |G_2\rangle-\epsilon |R\rangle$ with $\epsilon=(\Omega_2/2\Delta_2)\ll 1$. Our spin-1/2 system is then formed from $|\downarrow\rangle = |G_1\rangle$ and $|\uparrow\rangle= |D\rangle$. Here we assume that the positions of the atoms are fixed, which is an excellent approximation for ultracold systems over the timescales we consider.  The interaction Hamiltonian for such a gas of atoms projected onto the spin-1/2 states is, up to a irrelevant constant, 
\begin{equation}\label{spineq}
\hat{H} = (1/2)\sum_{j\neq k} \left[  (V_{jk}/4) \ \sigma^{z}_j \sigma^{z}_k  + (V_{jk}/2)\ \sigma^{z}_k \right]
\end{equation}
where the Pauli spin operators are $\sigma^{x}_k =  \left(|\downarrow\rangle_k \langle \uparrow|_k + \rm{h.c.} \right)$, $ \sigma^{y}_k = i \left(|\downarrow\rangle_k\langle \uparrow|_k - \rm{h.c.} \right)$, and $\sigma^{z}_k = \left(|\uparrow\rangle_k\langle \uparrow|_k - |\downarrow\rangle_k\langle \downarrow|_k \right)$. The interaction between dressed atoms $j$ and $k$ with inter-atomic distance $r_{jk}$ is  $V_{jk}=V(r_{jk})= \epsilon^4 C_6/(r^6_c+r^6_{jk})$. Here  $C_6$ is the van der Waals coefficient,  $r_c = |C_6/2\Delta_2|^{1/6}$ is the soft-core radius, and $V_0 = \epsilon^4(2\Delta_2)$ is the height of the dressed potential (we set $\hbar = 1$). This Hamiltonian is unique to Rydberg atoms in the sense that the coefficients of the Ising term and the single particle term are linked. This is because the interactions are dominated by the Rydberg states while interactions involving ground states are negligible.

Figure~\ref{setupfig}(c) shows the Ramsey protocols studied here. The first strong, resonant pulse, $(\Omega_1/2)(|G_1\rangle_k\langle D|_k + \text{h.c.})$, rotates the spins by $\theta$ around the $y$-axis. The wave function immediately after this pulse is
\begin{equation}
|\psi (t=0) \rangle =  \bigotimes_{k}  \big(\cos (\theta/2)|\downarrow\rangle_k + \sin (\theta/2)|\uparrow\rangle_k \big)
\end{equation}
where $\theta$ is proportional to the pulse area. During the Ramsey dark time $t$, the system evolves by Eq.~(1), developing correlations. After this time $t$, a second pulse rotates the spin component $\sigma^{\alpha=x,y}_k$ into the $z$ axis, where it can be measured as the population difference of $|D\rangle$ and $|G_1\rangle$. The incoherent emission from $|G_2\rangle$ is described by a master equation with jump operator $\sigma^-$ and rate $\gamma$ (see Fig.~\ref{setupfig}(c)). It is convenient to measure  and analyze the Ramsey contrast
\begin{eqnarray}
C(t) &=& |\sigma^+(t)|  = \sqrt{\langle\sigma^x(t)\rangle^2 + \langle\sigma^y(t)\rangle^2},\label{conphase}
\end{eqnarray}
and the phase, $\phi(t) = \arctan(\sigma^y(t)/\sigma^x(t))$. Here we defined $\sigma^{\alpha}  = \sum_k \sigma^{\alpha}_k$. We study the dynamics both with and without a spin-echo pulse, illustrated in Fig.~\ref{setupfig}(b). A $\pi$ spin-echo pulse around the $y$ axis (i.e. in phase with the first pulse) leaves the spin-model interactions invariant while removing single particle terms from the Hamiltonian Eq.~(\ref{spineq}), as well as any additional single-particle inhomogeneities in $\sigma^z_i$. The resulting dynamics are equivalent to evolution for time t {\it without} an echo but with effective Hamiltonian, $\hat{H}_{\rm echo} = (1/2) \sum_{j\neq k} (V_{jk}/4) \sigma^z_j \sigma^z_k$ as shown in the Supplementary Material~\cite{supp}. The  spin-echo has rather unusual effects in the Rydberg system as it effectively \textit{turns on} interactions between the $|G_1\rangle$ states, as illustrated in Fig.~\ref{setupfig}(c).

\begin{figure}[tb]
	\centering
	\includegraphics[width=\columnwidth]{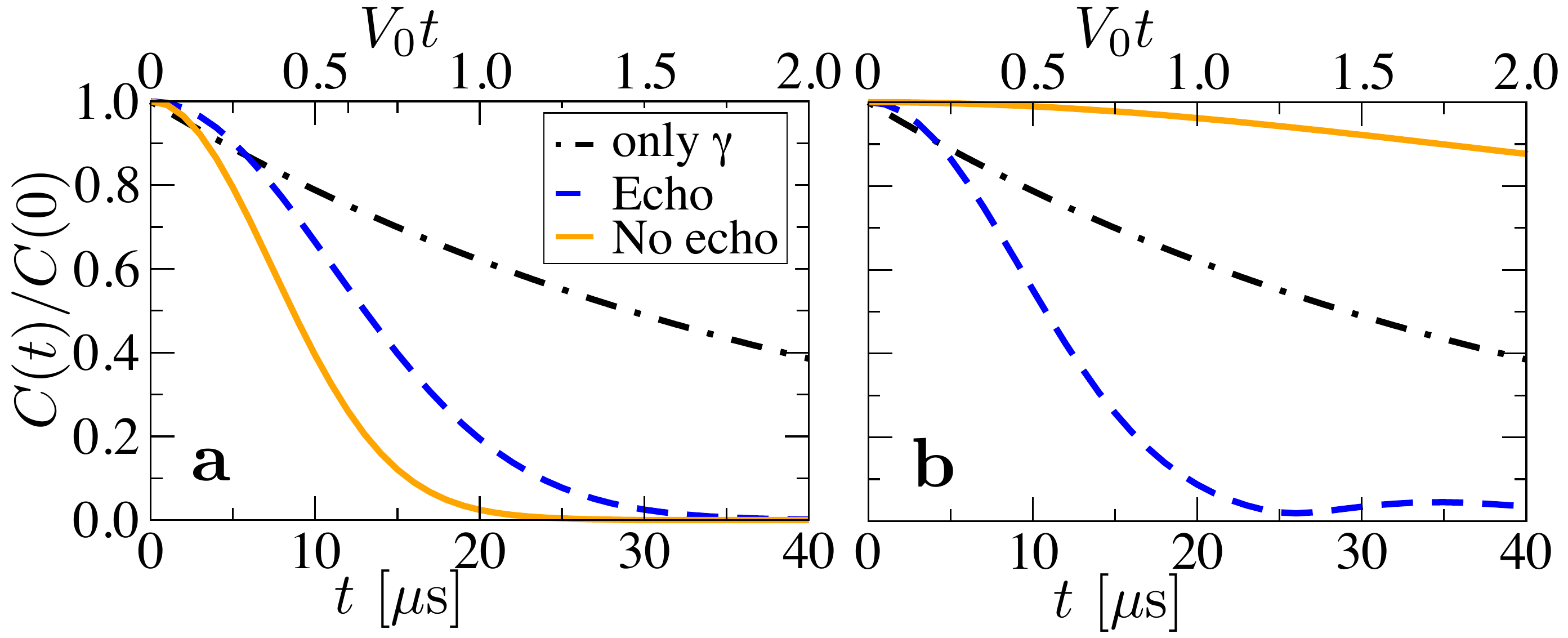}
	\caption{(color online) Ramsey contrast versus time for a gas of ground state Sr atoms with density $\rho=10^{12}/{\rm cm}^3$ dressed with $40$s triplet states with a dressing amplitude of $\epsilon=0.1$. Panels (a) and (b) show $\theta=\pi/2$ and $\pi/20$, respectively. Individual curves are for Ramsey with and without spin echo, compared to that of non-interacting atoms spontaneously emitting from only $\ket{G_2}$ with $\gamma = 21~\mu$s$^{-1}$.}
	\label{contrastdiss}
\end{figure}

{\it Calculating the Ramsey contrast and phase -} The spin dynamics of Eq.~(1) together with the non-trivial effects of the dissipation has been obtained by solving the corresponding master equation in Refs.~\cite{foss1,foss2} (generalizing Refs.~\cite{Emch,Raden,Kastner}), to obtain
\begin{equation}\label{fullana}
\langle\sigma^+(t)\rangle =
\sin\theta  e^{-2\gamma t}\sum_k \prod_{j\ne k} f(V_{jk}t) \\
\end{equation}
where 
\begin{eqnarray}\label{fullana2}
f(X) &=& e^{(i \beta X-\gamma t)/2} [\cos((X -i\gamma t)/2)\nonumber \\
\hspace{0.15in}
&&+\left((\gamma t -  i X\cos \theta )/2\right)~{\rm sinc}((X  -i\gamma t)/2)].
\end{eqnarray}
Here $\beta = 0$ is for echo while $\beta = 1$ is for no echo dynamics. The function $f(X)$ often simplifies; for example, $f(X)=\cos(X/2)$ for $\theta=\pi/2$ and $\gamma=0$. Another relevant decoherence is a dephasing of the $|G_2\rangle$ level at rate $\gamma_d$ as could result from laser noise; this can be included simply by multiplying $\langle\sigma^+\rangle$ by $e^{-\gamma_d t}$. We neglect such noise as it depends on the details of the experiment, and often is negligible compared to $\gamma$. 

Although Eq.~(\ref{conphase}) allows us to calculate the dynamics once we know the atom positions, it often is impossible to measure the positions of all of the atoms. However, in a large enough system, the dynamics is expected to ``self-average": $C(t)$ for a single configuration in a large system is equal to its average over all configurations, and therefore it is independent of the specific configuration. We model the atoms to be independently distributed with a uniform density $\rho$ for simplicity. The assumption is quantitatively justified for an initially weakly interacting, not-too-degenerate gas. Rather remarkably, we are able to analytically perform this disorder average in the thermodynamic limit: we find that Eq.~(\ref{fullana}) simplifies to evaluating a one dimensional integral (see Supplementary Material \cite{supp}),
\begin{equation}\label{simpleana}
\langle\sigma^+(t)\rangle = \exp\left(-\rho \int\!4\pi r^2 dr \, \left[1-f(V(r)t) \right] \right).
\end{equation}

 \begin{figure}[tb]
 	\centering
 	\includegraphics[width=\columnwidth]{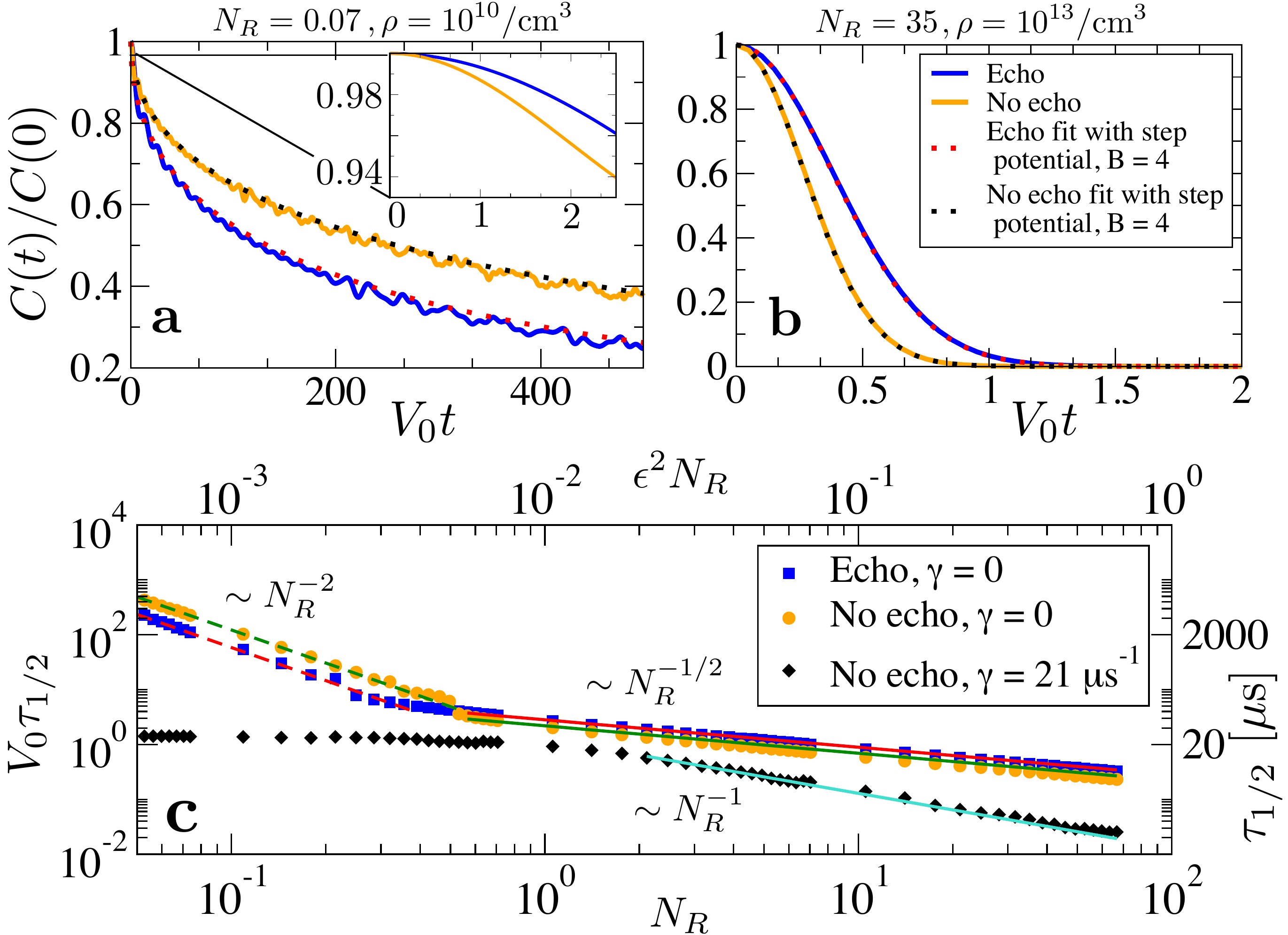}
 	\caption{ (color online) (a-b) Contrast dynamics obtained from Eq.~(\ref{ct}) takes a dramatically different shape in the $N_R  \ll 1$ and $N_R \gg 1$ limits.  Interestingly, for $N_R \ll 1$ the contrast dynamics is faster with echo than without, except at exceptionally short times (inset).  The dashed lines are the analytic predictions of Eq.~(6) for these limits. (c) Characteristic timescale for the dynamics as a function of $N_R$. The top \textit{x}-axis represents the average number of Rydberg excitations inside $r_c$.}
 	\label{densityscaling}
 \end{figure}

{\it Characterizing Rydberg-dressed interactions using Ramsey scheme -} In the absence of dissipation, Eq.~(\ref{simpleana}) implies that the dimensionless parameters, $N_R = 4\pi \rho r_c^3/3$ and $V_0  t$ encapsulate the dependence of the dynamics  on density $\rho$, van der Waals coefficient $C_6$, and optical parameters ($\Omega_2$, $\Delta_2$). Thus, we present the contrast in Figs.~\ref{contrastdiss}-~\ref{densityscaling} as a function $C(N_R,V_0t)$, from which one can easily extract the dynamics for any experiment by using the appropriate $N_R$ and $V_0$. To show typical scales, we also show the dynamics for typical Sr experimental parameters~\cite{Ye} with $C_6$ coefficients from Ref.~\cite{Vaillant}. 

Figure~\ref{contrastdiss}(a) demonstrates that the interaction-driven dynamics occurs on an experimentally favorable timescale, both with and without a spin echo. For example, in Sr it is substantially faster than the spontaneous emission rate $\gamma=21\mu$s$^{-1}$ of the $\ket{G_2}=\ket{{}^3P_1}$ state. Typically the dynamics of the  non-echo is faster than echo due to single particle inhomogeneities. Interestingly, due to the unique structure of the Rydberg interactions, this naive intuition sometimes fails, as seen in Fig.~\ref{contrastdiss}(b) for $\theta\ll 1$. In this case, the spin-echo pulse increases the effective interactions since it converts initially non-interacting $\ket{G_1}$ atoms into strongly interacting $\ket{D}$ states. 

Figure~\ref{densityscaling} shows that the contrast is sensitive to the shape of the potential, with striking differences in the low-density ($N_R\ll1 $) and high-density ($N_R\gg 1$) limits. These differences arise because for $N_R\ll1$ the dynamics probes the $1/r^6$ interaction tail, while for $N_R\gg1$ it probes the core. In these limits, the disorder-averaged contrast of Eq.~(\ref{simpleana}) simplifies to
\begin{equation}\label{ct}
C(N_R,V_0 t) =\begin{cases}
e^{-A N_R\sqrt{V_0 t}}~&(N_R \ll 1) \\
e^{-B N_R (1-\cos^{\beta +1} (V_0 t)/2)}~&(N_R \gg 1) 
\end{cases}
\end{equation}
with $A = \sqrt{\pi}/2^{1+\beta/2}$ where $\beta=0$ is for echo dynamics and $\beta = 1$  for non-echo dynamics. For $N_R\ll1$, this contrast is the exact solution of Eq.~(\ref{simpleana}), and is non-analytic at $t=0$; thus it is beyond all orders of perturbation theory. For $N_R\gg 1$, Eq.~(\ref{ct}) approximates $V(r)\approx V_0 H(r_c-r)$, where $H(x)$ is the Heaviside function. As shown in Fig.~\ref{densityscaling}(b), this simple model reproduces the exact contrast up to an overall shift of the timescale: the naive $B=1$ of the hard core potential is replaced by $B=4.0$ for the shown value of $N_R$. The scaling of the characteristic time $\tau_{1/2}$, defined as $C(N_R,V_0\tau_{1/2}) = C(N_R,0)/2$, follows directly from Eq.~(\ref{simpleana}) as
\begin{equation}
\tau_{1/2} \propto (N_R)^{\alpha}/V_0 \propto \Omega^{-4}_2~\Delta^{3-\alpha/2}_2~C_6^{\alpha/2} \rho^{\alpha}
\end{equation} 
where $\alpha = -2$ for $N_R \ll 1$, while $\alpha = -1/2$ for $N_R\gg 1$. This scaling is confirmed in Fig.~\ref{densityscaling}(c), where we also include results for Sr with dissipation. Fig.~\ref{densityscaling}(c)'s top axis shows  $\epsilon^2 N_R$,  which must be small in order for Eq.~(\ref{spineq}) be valid. Although Fig.~\ref{densityscaling}(c) shows that the dissipation substantially increases the rate of contrast decay, this is not due to to naive single-particle decay, which occurs on the slow 21$\mu$s timescale, as shown in Fig.~\ref{setupfig}. Rather, this is due to the back action of the emission events on the rest of the spins through the interaction, which induces strong correlations; this is analogous to the feedback effect explained in Ref.~\cite{foss1}.

We note that losses due to ionization or molecular resonances are neglected in Eq.~(\ref{fullana}).  One way to avoid resonances is to confine atoms in a lattice with appropriate lattice spacings~\cite{Vermersch, Derevianko}. Recently it has also been shown that for large number of atoms, losses to other dipole allowed Rydberg states can be significant \cite{Goldschmidt, Camargo}. All of these are beyond the scope of this paper, but often should be relevant only on timescales beyond those of interest here.

\begin{figure}[tb]
	\centering
	\includegraphics[width=\columnwidth]{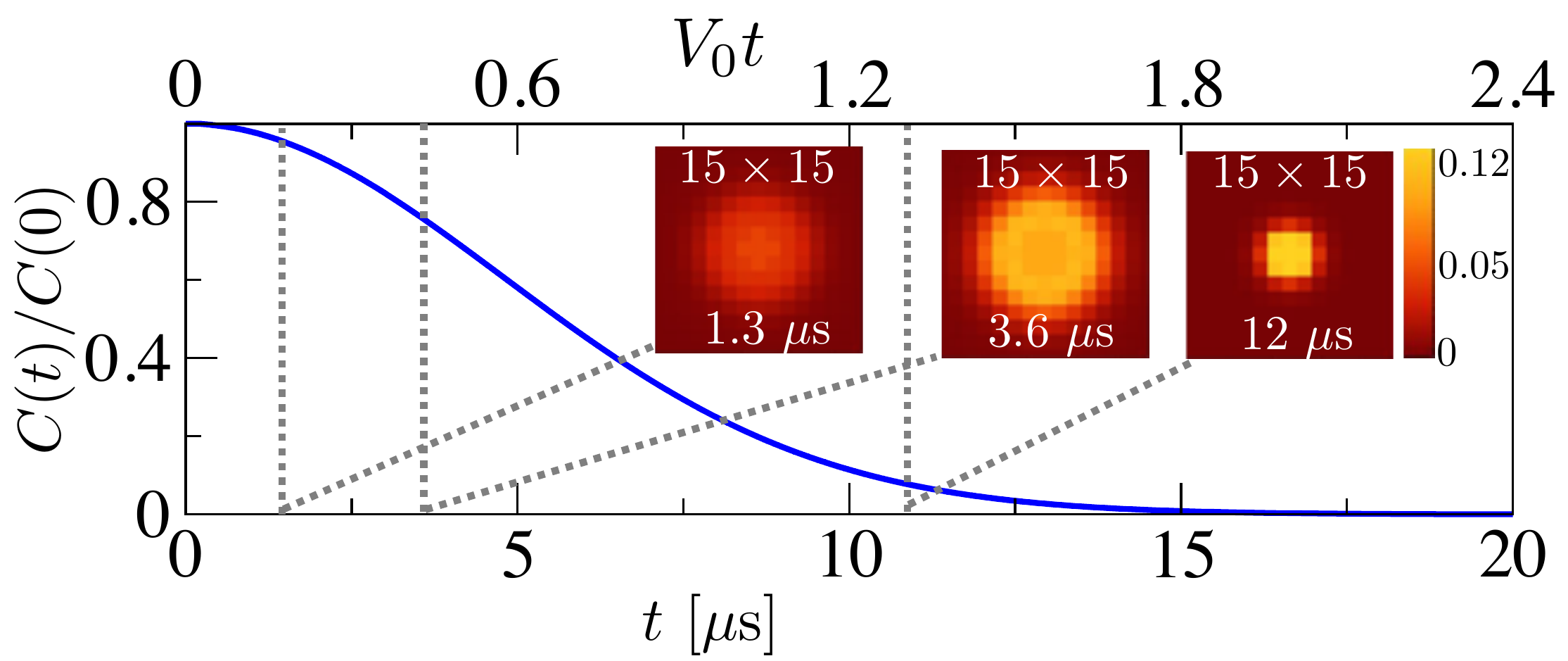}
	\caption{(color online) Contrast and correlation dynamics in a unit-filled $15 \times 15$ square lattice. Dressing parameters are the same as in Fig.~\ref{contrastdiss}. The lattice spacing is $a=0.5~\mu$m and $r_c \simeq 2a$. The connected correlation functions $\mathcal{G}$ between the center atom and the surrounding atoms is  shown in the insets.}
	\label{g2fig}
\end{figure}

{\it Correlations using Rydberg-dressed interactions in a lattice -} Our calculations so far have been for a gas where both the excitation and measurement are done collectively on the whole system; this is the typical and most straightforward case in current experiments. However, it is also possible to spatially resolve $\langle S^x_i \rangle$~\cite{Schwarzkopf, McQuillen, Gunter}. The extreme limit of this capability is the single atom resolution that has recently been achieved in  microscope experiments~\cite{Kuhr, Schauss1, Schauss2}. The images in these experiments reveal not only $\langle S^x_i \rangle$, but also correlations such as $\langle S^x_i S^x_j \rangle$. 

Motivated by these ongoing experiments, we study the Ramsey dynamics on a unit-filled two-dimensional square lattice. Fig.~\ref{g2fig} shows the $C(t)$ and snapshots of the connected correlation~\cite{foss1, foss2} $\mathcal{G}(i,j)= \langle S^x_i S^x_j \rangle - \langle S^x_i \rangle \langle S^x_j \rangle$. We observe that the shape and timescale of the contrast dynamics is similar to the $N_R\gg 1$ limit found in the gas, and that the decay of the contrast is associated with a growth of strong spin correlations within a radius $r_c$.

\begin{figure}[tb]
	\centering
	\includegraphics[scale=0.4]{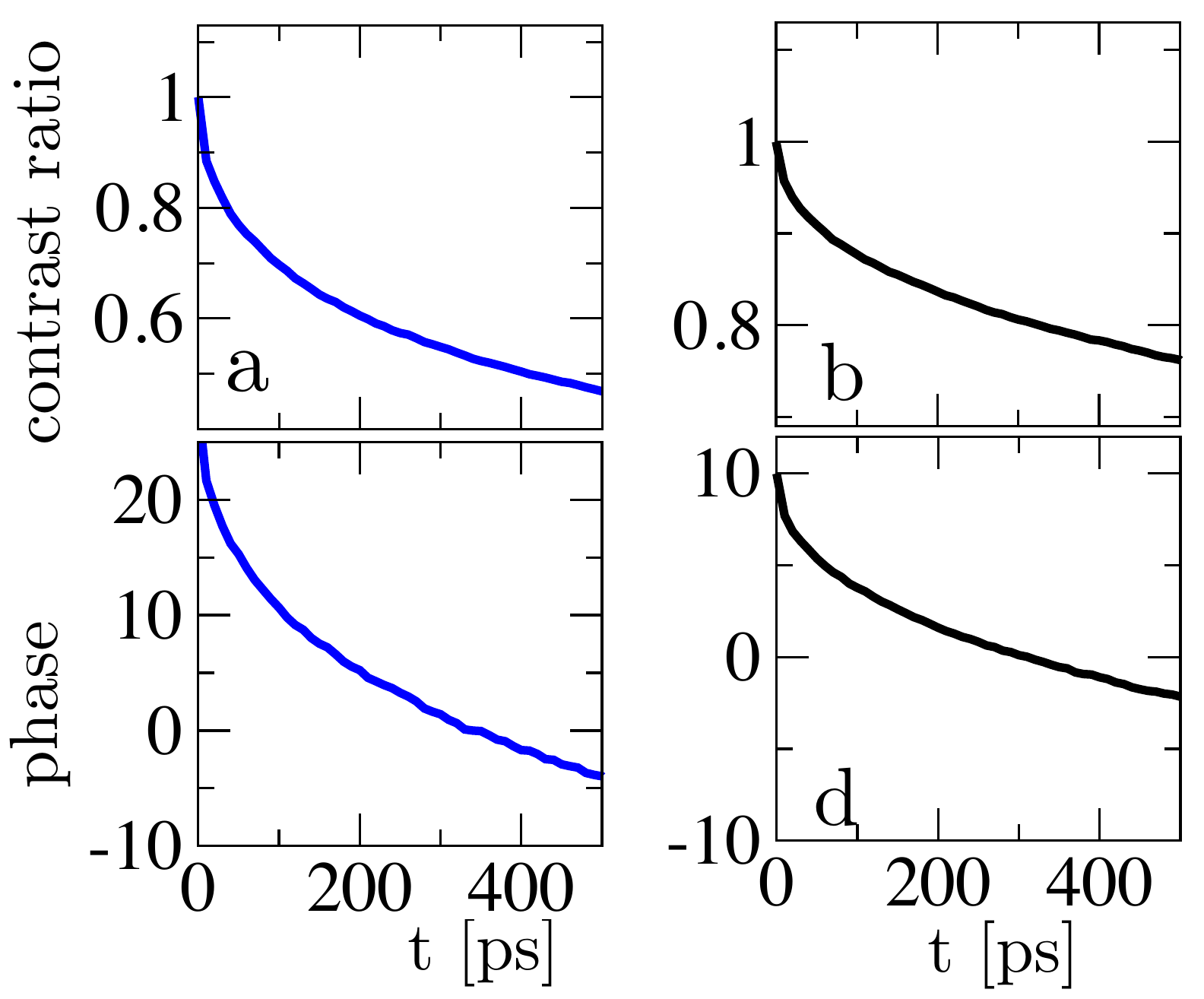}
	\caption{(color online)  Calculated contrast decay and phase shift for bare Rydberg atoms, which agrees with experimental measurements (which will be shown pending publication of Ref. \cite{Takei}). Left and right columns are for 3.1\% and 1.2\% Rydberg fraction, respectively. (a,b) are the ratio of the contrast at $\rho=1.3\times 10^{12}/{\rm cm}^3$ to the contrast at $4\times 10^{10}/{\rm cm}^3$.  (c,d) are the corresponding Ramsey phase shifted at $t=0$ as in the analysis in Ref.~\cite{Takei}.}
	\label{Rbpumpprobe}
\end{figure}

{\it Characterizing bare Rydberg interactions using ultrafast lasers -} To exemplify the generality of our calculations we consider the recently performed ultrafast Ramsey experiment with bare Rydberg states of Rb atoms~\cite{Takei}. This can be viewed as a special case of Rydberg-dressed atoms with $\epsilon=1$ and $r_c=0$, so one obtains a pure van der Waals potential. Typically, due to strong blockade effects, it is hard to excite atoms to superpositions of ground and bare Rydberg states at a sufficiently large density. Ref.~\cite{Takei} overcomes this difficulty using strong, ultrafast lasers to couple atoms from their ground state $|G\rangle$ to the bare Rydberg state $|R\rangle = 42D_{5/2}$. We calculate the contrast and phase for this experiment using the $C_6$ coefficients from Ref.~\cite{Reinhard}. Fig.~\ref{Rbpumpprobe} shows our theory, which quantitatively agrees with experiment without any fitting and also coincides with the calculations in \cite{Takei}.This agreement confirms that the effects of Zeeman degeneracies ~\cite{Walker} can be neglected and that the low fraction of Rydberg excitation suppresses any resonances. This provides a compelling proof of principle for this detection method in future ultracold  Rydberg-dressed experiments. 

{\it Conclusion -} Although previous experiments on coherent many-body excitation with Rydberg atoms have demonstrated strong correlations, they have not established their \textit{non-classicality}. Ramsey spectroscopy directly creates and probes the superpositions necessary for quantum correlations. We have shown that this dynamics happens on experimentally favorable timescales. The exact analytic solutions available for the dynamics will allow for systematic comparison between experiment and theory, despite the strongly correlated dynamics. (To emphasize the strong correlations, note that for $\theta=\pi/2$, the dynamics is completely beyond mean-field theory, which predicts a time-independent spin-echo contrast). We showed, as a first step in this direction, that our theory quantitatively agrees with recent ultrafast measurements of bare Rydberg atoms, a remarkable demonstration of the universality of the dynamics over six orders of magnitude, from $\mu$s to ps. We revealed that the contrast dynamics is sensitive to the shape of the interaction potential, as well as to density, principal quantum number, and dressing laser properties. Striking dynamics emerges for $N_R\ll 1$: The contrast displays a non-perturbative short time non-analyticity, $e^{-\sqrt{t/\tau_0}}$, due to the $1/r^6$ character of Rydberg interactions. Further interesting dynamical phenomena result from the unique nature of Rydberg interactions. For example, the spin echo pulse enhances the rate of contrast dynamics for small excitation fraction $\theta$ or low densities.
 
In the future, Ramsey experiments with local pulses applied at various phase delays would allow measurement of the quantumness in the correlations, and would pave the way for studying interesting many-body non-equilibrium spin physics. Another interesting direction is to simply add a transverse field. This prevents exact solution and the resulting rich many-body dynamics will be an interesting testbed for theoretical methods~\cite{Schachenmayer,Schachenmayer2,Schachenmayer3,Pucci}.

\textit{Acknowledgements -} We thank M. Foss-Feig, M. Wall, B. DeSalvo, B. Dunning, T. Pohl,  J. Zeiher and C. Gross for useful conversations. We also acknowledge N. Takei, C. Sommer, C. Genes, G. Pupillo, M.~Weidem{\"u}ller  and K. Ohmori for discussions on the ultra-fast experiments. Part of this work was performed at the Aspen Center for Physics, which is supported by National Science Foundation grant PHY-1066293. This work was supported with funds from the Welch foundation, Grant No. C-1872.

\clearpage

\onecolumngrid

\hrulefill
\begin{center}
	\Large\textbf{Supplemental Material for "Accessing Rydberg-dressed interactions using many-body Ramsey dynamics"}
\end{center}

\section{Distinguishing Rydberg spin dynamics between non-echo and echo}

In the non-echo dynamics protocol, the time evolution operator for the Ramsey dark period is given by $U_{\rm non-echo} = e^{-i\hat{H}t}$, where $\hat{H}$ is in Eq.~(1). For the spin echo dynamics, the time evolution operator is given by $U_{\rm echo} = e^{-i\hat{H}t/2}R_{\pi}e^{-i\hat{H}t/2}$ where $R_{\pi}$ is the $\pi-$pulse evolution. The combined effect of $R_{\pi} e^{-i\hat{H}t/2}$ is that the system evolves under a unitary transformation given by $e^{-i\hat{H}_R t/2}$, where $\hat{H}_R$ is same as $\hat{H}$ but with $\sigma^z_j \rightarrow -\sigma^z_j$ :
\begin{equation}
\hat{H_R} = (1/2)\sum_{j\neq k} \left[  (V_{jk}/4) \ \sigma^{z}_j \sigma^{z}_k  - (V_{jk}/2)\ \sigma^{z}_k \right] ,
\end{equation}
so $U_{\rm echo} =e^{-i \hat{H} t/2} e^{-i \hat{H}_R t/2}$. Since $\hat{H}$ and $\hat{H_R}$ commute, the resulting time evolution operator  is $U_{\rm echo} =e^{-i \hat{H}_{\rm echo}t}$ where $\hat{H}_{\rm echo} = (\hat{H}+\hat{H}_R)/2$. Physically this implies that interactions between atoms initially in $|G_1\rangle$ states start interacting in the second half of the spin echo dynamics. Additional contributions  arising from interactions with atoms in $|G_1\rangle$ and $|D\rangle$ states are negligible.

\section{Analytic expression of contrast for a uniformly distributed gas of atoms}

For the derivation of Eq.~(6), consider a uniformly distributed gas of $N$ atoms with independent positions labeled by $j,k$. Since the particles are independent, the probability distribution in space factors as $P(\mathbf{r}_1, \mathbf{r}_2, \hdots \mathbf{r}_N) = P(\mathbf{r}_1) P(\mathbf{r}_1) \hdots P(\mathbf{r}_N)$. For a uniform distribution, we have $P(\mathbf{r}_j) = \rho/N$ where $\rho$ is the uniform density of the gas. Thus averaging Eq.~(4) in the main text, we  have 
\begin{eqnarray}
\langle \sigma^+_k(t) \rangle  &=& \int \! d\mathbf{r}_1 \,\! d\mathbf{r}_2 \,\hdots \, d\mathbf{r}_{N-1} \, \left(\frac{\rho}{N-1}  \right)^{N-1} \left[\prod_{j\ne k} f(V(r_{jk}) t)\right]  \nonumber \\ 
&=&  \left(\frac{\rho}{N-1}\right)^{N-1}  \prod_{j\ne k} \left[\int \! d\mathbf{r}_j \, f(V(r_{jk}) t)\right] \nonumber \\ 
&=& \left[\left(\frac{\rho}{N-1}\right)\int \! d\mathbf{r}_j \, f(V(r_{jk}) t) \right]^{N-1} \ .
\end{eqnarray}
The remaining single integral diverges with the system volume, and thus it is convenient to define the finite integral
\begin{equation}
\mathcal{I} = \rho \int\!4\pi r^2dr\,[1-f(V(r) t)] ,
\end{equation}
in terms of which the contrast is
\begin{equation}
\langle \sigma^+(t) \rangle  = \left[1-\frac{\mathcal{I}}{N} \right]^{N-1} \ .
\end{equation}
In the thermodynamic limit where $N\rightarrow \infty$, the above expression is Eq.~(6) in the main text. Note that this simplified expression depends on the function $f$, which in turn depends on the exact dynamics (echo, non-echo, tipping angle, dissipation). Our results coincide with those of \cite{Takei} in the special cases they calculate.

\end{document}